\documentclass[nobibnotes,superscriptaddress,nofootinbib,twocolumn,floatfix]{revtex4}

\usepackage{amsfonts}
\usepackage{amssymb}
\usepackage{graphicx}
\usepackage{amsmath}
\usepackage{float}

\begin{document}
\title{Quantum magnetic collapse of a partially bosonized $npe$-gas: Implications for astrophysical jets}
\author{R. Gonz\'alez Felipe}
\email{ricardo.felipe@tecnico.ulisboa.pt}
\affiliation{ISEL - Instituto Superior de Engenharia de Lisboa, Instituto Polit\'ecnico de Lisboa, Rua Conselheiro Em\'{\i}dio Navarro, 1959-007 Lisboa, Portugal\\
CFTP - Centro de F\'{\i}sica Te\'{o}rica de Part\'{\i}culas, Instituto Superior T\'{e}cnico,
Universidade de Lisboa, Avenida Rovisco Pais, 1049-001 Lisboa, Portugal\\}
\author{A. P\'erez Mart\'{\i}nez}
\email{aurora@icimaf.cu}
\affiliation{Instituto de Cibern\'{e}tica, Matem\'{a}tica y F\'{\i}sica (ICIMAF), \\
 Calle E esq a 15 Vedado 10400 La Habana Cuba}
\author{H. P\'erez Rojas}
\email{hugo@icimaf.cu}
\affiliation {Instituto de Cibern\'{e}tica, Matem\'{a}tica y F\'{\i}sica (ICIMAF), \\
 Calle E esq a 15 Vedado 10400 La Habana Cuba}
\author{G. Quintero Angulo}
\email{gquintero@fisica.uh.cu}
\affiliation{Facultad de F{\'i}sica, Universidad de la Habana,\\ San L{\'a}zaro y L, Vedado, La Habana 10400, Cuba}

\begin{abstract}
We study a possible mechanism for astrophysical jet production from a neutron star composed by a partially bosonized $npe$-gas. We obtain that the expulsion of a stable stream of matter might be triggered by the quantum magnetic collapse of one or various components of the gas, while its collimation is due to the formation of a strong self-generated magnetic field.
\end{abstract}


\maketitle

\section{Introduction}
\label{sec1}
Astrophysics is living a golden period after the appearance of new tools to explore the Universe and, in particular, among its more interesting components, the neutron stars (NS). The direct detection of gravitational waves in the past four years~\cite{Abbott:2016blz, TheLIGOScientific:2017qsa,GBM:2017lvd} has opened the door to listening to our Universe. At the same time, the progressive improvement of the observational techniques as well as the launch of new spatial observatories like the Neutron Star Interior Composition Explorer (NICER) Mission\footnote{https://heasarc.gsfc.nasa.gov/docs/nicer/} complement this new scenario and will allow to obtain more accurate constraints on NS observables. In the years to come, the results of these observations and their interpretations will contribute to the challenge of understanding the physics of the matter in the interior of neutron stars, the densest objects in the Universe, as well as some others exotic phenomena like astrophysical jets.

Astrophysical jets are streams of collimated matter that might be ejected by several objects (stars, protostars, protoplanetary nebulae, compact objects, galaxies, etc.)~\cite{deGouveiaDalPino:2004jy,Blandford:2018iot}. Depending on the source, jets may have different size, scale and velocities. Nevertheless, they all bear two outstanding features that make them different from any other object in the Universe: their elongated form and the fact that the matter composing them moves away from the source without dispersing. The elongated form of jets is particularly intriguing, if one recalls that most astronomical objects are spherical or oblate shaped, due to the combination of the angular momentum conservation and the central symmetry of the gravitational force. That is the reason why, regardless of their origins, it is believed that all jets are produced and maintained by similar mechanisms. Although the physics behind jet production and maintenance is still under debate, the general consensus is that magnetic fields play an important role~\cite{deGouveiaDalPino:2005xn,Pino:2010qs,Blandford:2018iot}.

Theoretical studies on jet formation seek for space-time metrics that allow to construct gravitomagnetic jets~\cite{Tucker:2016wvt,Bini:2007zzb,Chicone:2010aa,Poirier:2015hga,Bini:2017uax}, as well as for mechanisms to explain their origins~\cite{Massi:2010gr}. Laboratory experiments are also being conducted with the aim of reproducing jets~\cite{Lebedev:2001fa,Hartigan}. Our current understanding of relativistic jet formation is mainly based on 3D simulations performed in the context of general relativistic magneto-hydrodynamics~\cite{Meier2012,Nakamura:2018htq}. Some attempts to simulate numerically astrophysical jets, for instance, the M87 from giant elliptical galaxy (a.k.a Virgo A, NGV4486, and 3C274), have also been made in the framework of specific gravity theories~\cite{Armengol:2016xhu}.

In this paper, we propose a possible mechanism for astrophysical jet formation and maintenance based on two properties of magnetized quantum gases, namely, quantum magnetic collapse and self-magnetization~\cite{2000PhRvL..84.5261C,Felipe:2002wt,Aurora2003EPJC,Felipe:2008cm,Ferrer:2015wca,Quintero2017PRC}. In most dense and cold quantum gases, the presence of a magnetic field produces a decrease in the pressure in the direction perpendicular to the magnetic axis. If the magnetic field strength and densities are high enough, the perpendicular pressure component may become negative, pushing the particles toward the magnetic axis and producing an instability that results in a cigar shaped structure. Such an instability, known as quantum magnetic collapse~\cite{2000PhRvL..84.5261C}, might perfectly be the ultimate cause that originates the jet. On the other hand, the huge magnetic fields required for the occurrence of the magnetic collapse might also result in the phenomenon of self-magnetization, produced by the alignment of the magnetic moment of the particles that compose the jet.

To explore the possibility that both effects, quantum magnetic collapse and  self-magnetization, can actually be behind jet formation, we shall assume that the jet is originated inside a neutron star. In addition, we consider a simple scenario of a NS composed by an ideal gas of neutrons, protons and electrons in which a fraction of the nucleons is paired forming composite bosons~\cite{Chavanis,Angulo:2018url}. We refer to such a mixture of gases as a partially bosonized $npe$-gas.

The paper is organized as follows. In Sec.~\ref{sec2}, we study the magnetic collapse of a partially bosonized $npe$-gas and a heuristic argument in favor of this physical phenomenon is given. Section~\ref{sec3} is devoted to the generation of the magnetic field inside the NS, and the study of the equations of state (EoS) of the collimated matter. Finally, our concluding remarks are given in Sec.~\ref{sec4}. A brief summary of the EoS of each $npe$-gas component is presented in Appendix~\ref{appA}.

\section{Partially bosonized $npe$-gas under the action of a magnetic field}
\label{sec2}

We consider a neutron star composed by a mixture of free neutrons $n$, protons $p$, and electrons $e$  - gas \textit{npe} -, and paired (bosonized) neutrons $nn$ and protons $pp$. We assume that there are no interactions between these  particles, i.e. all species are considered as ideal relativistic gases. We also assume  that particles are under the action of a locally uniform and constant magnetic field, directed along the $z$ axis, i.e., $\textbf{B}=(0,0,B)$.

The thermodynamic description of each gas in the mixture is given in Appendix~\ref{appA}, where the EoS are obtained starting from the general expression of the thermodynamical potential
\begin{align}
&\Omega_{i}(B,T,\mu)=\frac{T}{4\pi}\sum_{s}dp_{\parallel}d^2p_{\perp}\nonumber\\
&\times\ln\left[\left(1+e^{-(\varepsilon_i-\mu_i)/T}\right)\left(1+e^{-(\varepsilon_i+\mu_i)/T}\right)\right],
\end{align}
where $T$ is the temperature, $s$ are the spin projections, and $p_{\parallel}$ and $p_{\perp}$ are the particle momentum components along and perpendicular to the magnetic field direction, respectively; $\varepsilon_i$ is the energy spectrum, $\mu_i$ is the chemical potential, and the index $i$ denotes the particle species, i.e. $i= e, p, n, nn, pp$. In what follows, we will work in the $T=0$ limit for fermions, while keeping temperature effects for bosons. The difference in the treatment is justified by the fact that, although in stellar conditions both kind of gases can be considered strongly degenerated, bosons remains very sensitive to temperature effects due to the drastic changes on the gas entropy and pressure caused by Bose-Einstein condensation.

The effects of the magnetic field are included in the thermodynamical potential through the spectrum of the particles. The spectrum of the unpaired and paired nucleons in the presence of a magnetic field $B$ are given by the expressions
\begin{equation}\label{fermionsGS}
\varepsilon_f =\left\{
\begin{array}{lr}
\sqrt{p_{\parallel}^2+ 2 q Bl+m ^ 2}, &\text{charged fermion,} \\[3mm]
\sqrt{p_{\parallel}+ \sqrt{p_{\perp}^2+m^2} -\kappa B}, &\text{neutral fermion,}  \\
\end{array}\right.
\end{equation}
and
\begin{equation} \label{bosonsGS}
\varepsilon_b = \left\{
\begin {array} {lr}
\sqrt{p_{\parallel}^2 - q B l+ m ^ 2}, & \text{charged scalar boson,}\\[3mm]
\sqrt{p_{\parallel}^2+ \sqrt{p_{\perp}^2+m ^ 2}-\kappa B}, & \text{vector neutral boson,}
\end{array}\right.\\[2mm]
\end{equation}
where $m$ denotes the mass of the particles, $q$ is the electric charge, and $\kappa$ the magnetic moment. In the above equations, the fermion spectrum is obtained by solving the Dirac equation for charged and neutral particles~\cite{Felipe:2002wt,Aurora2003EPJC}, while the boson spectrum is derived from the Klein-Gordon equation~\cite{ROJAS1996148} in the case of scalar particles, and from the Proca equation in the case of the vectorial ones~\cite{Quintero2017PRC}.

Once $\Omega_i$ is determined, the EoS of each gas is obtained through the usual expressions for magnetized gases~\cite{Ferrer:2015wca}, namely,
\begin{align}
P_{i\, \parallel}&=-\Omega_i, \quad P_{i\, \perp}=-\Omega_i-\mathcal M_i B,\nonumber\\[2mm] 
E_i&=-\Omega_i+\mu_i N_i+T S_i,\\[2mm]
N_i&=-\frac{\partial\Omega_i}{\partial \mu_i}, \quad
\mathcal M_i=-\frac{\partial\Omega_i}{\partial B}, \quad S_i=-\frac{\partial\Omega_i}{\partial T},\nonumber
\end{align}
where $P_{i\,\parallel}$ and $P_{i\,\perp}$ are the pressures of the gas along and perpendicular to the magnetic field direction, respectively; $\mathcal M_i$ is the magnetization, $E_i$ is the internal energy, $N_i$ is the particle number density, and $S_i$ is the entropy. Note that the above EoS contains the anisotropy in the pressures ($P_{i\,\parallel} \neq P_{i\,\perp}$) induced in the system by the presence of the magnetic field.

The total thermodynamic quantities of the gas mixture, i.e., the pressures $P^T_{\parallel}$ and $P_{\perp}^T$, the magnetization $\mathcal M^{T}$ and the energy $E^T$, are then calculated as
\begin{align}\label{edenpe}
P^T_{\parallel} &= \sum_{i} P_{i\,\parallel},\quad
P^T_{\perp} = \sum_{i} P_{i\,\perp},\nonumber\\
\mathcal M^T &= \sum_{i} \mathcal \mathcal M_i,\\
E^T &= \sum_{i} E_i.\nonumber
\end{align}

We will also consider that the partially bosonized $npe$-gas is in stellar equilibrium, so that the following conditions are satisfied:
\begin{align}\label{estelarcond}
N_{n}^T+N_{p}^T&=N_B,\quad \text{baryon number conservation},\nonumber\\
N_p^T&=N_e,\quad \text{charge neutrality},\\
\mu_p+\mu_n &=\mu_e, \quad \text{$\beta$-equilibrium},\nonumber
\end{align}
where $N_p^T = N_p+2N_{pp}, N_n^T = N_n+2N_{nn}$ are the total number of protons and neutrons, respectively; $N_B$ is the baryonic number density; $x_n=2 N_{nn}/N_n^T$ and $x_p=2 N_{pp}/N_p^T$ correspond to the fractions of total bosonized neutrons and protons, respectively.

The fractions of paired nucleons, $x_n$ and $x_p$, depend in general on the density and the temperature. However, such a dependence is presently unknown, since we are unable to reproduce the astrophysical conditions in terrestrial laboratories. Therefore, hereafter we will take them as free parameters that are set externally to the system. For simplicity, in our numerical calculations, we shall assume $x_n = x_p=0.5$.

Under the stellar conditions given in Eqs.~(\ref{estelarcond}) and neglecting the effect of bosonization in the $\beta$-equilibrium condition, the fraction of protons (and electrons) for a given neutron density reads~\cite{Weinberg,Shapiro}
\begin{equation}\label{protonfraction}
\frac{N_{p}^T}{N_n^T} = \frac{1}{8} \left[\frac{1 + 4 Q/(m_n y_n^2) + 4 (Q^2 - m_e^2)/(m_n^2 y_n^4)}{1+1/y_n^2} \right]^{3/2},\\[3mm]
\end{equation}
where $Q = m_n-m_p$, $y_n = p_F/m_n$ and $p_F=({\mu}^2-\varepsilon_{n}^2)^{1/2}$ is the Fermi momentum of neutrons.

\begin{figure}[t]
	\centering
	\includegraphics[width=1\linewidth]{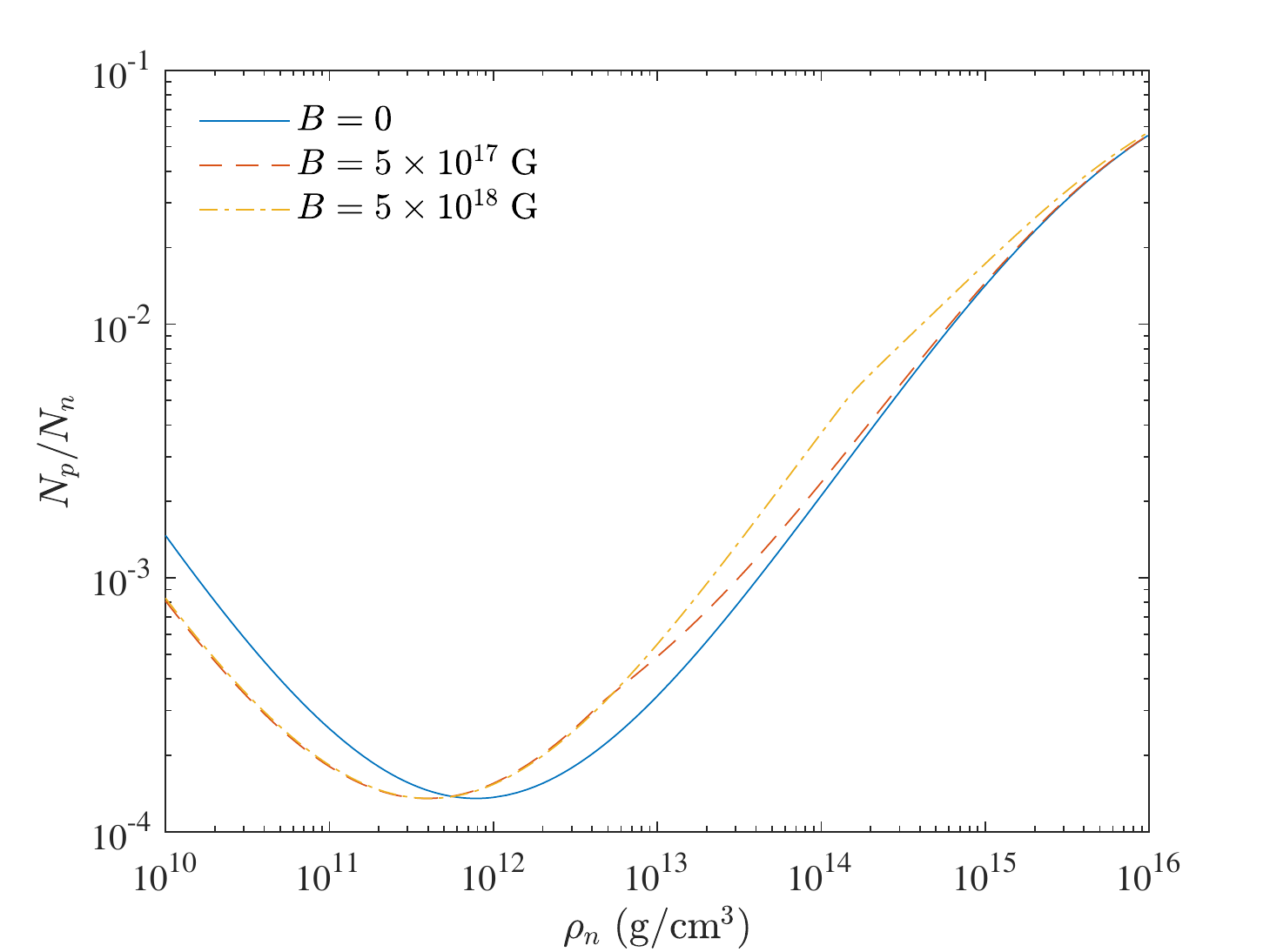}
	\caption{\label{protonelectronfraction} Proton fraction as a function of the neutron mass density for $B=0$, $B=5\times 10^{17}$~G, and $B=5\times 10^{18}$~G. The curves shift to left as the magnetic field intensity increases.}
\end{figure}

In the presence of the magnetic field, the proton and electron fractions depend on its intensity $B$ through ${p_F}$. This dependence can be seen in Fig.~\ref{protonelectronfraction}, where the proton fraction given in Eq.~\eqref{protonfraction} is plotted as a function of the neutron mass density $\rho_n = m_n N_n$ for different values of $B$. Increasing the magnetic field shifts the minimum of the proton fraction to the left and, consequently, to the region of lower densities. Also, an increase of the magnetic field increases the fraction of protons (and electrons) inside the star in the high density region.

\subsection{Quantum magnetic collapse}
\label{subsec2a}

The magnetic collapse of quantum gases have been previously discussed as a mechanism that provokes one of the pressure components to become negative~\cite{2000PhRvL..84.5261C,Quintero2017PRC}. If the perpendicular pressure becomes less than zero, this collapse would be transverse to the magnetic axis and the particles of the gas would be pushed toward it. This gives rise to the formation of an elongated and axisymmetric structure with a cigar shape, supporting the idea that the ejection of mass out of the star could be due to the transverse magnetic collapse of one or more  quantum gas species. 

\begin{figure}[t]
\centering
\includegraphics[width=1\linewidth]{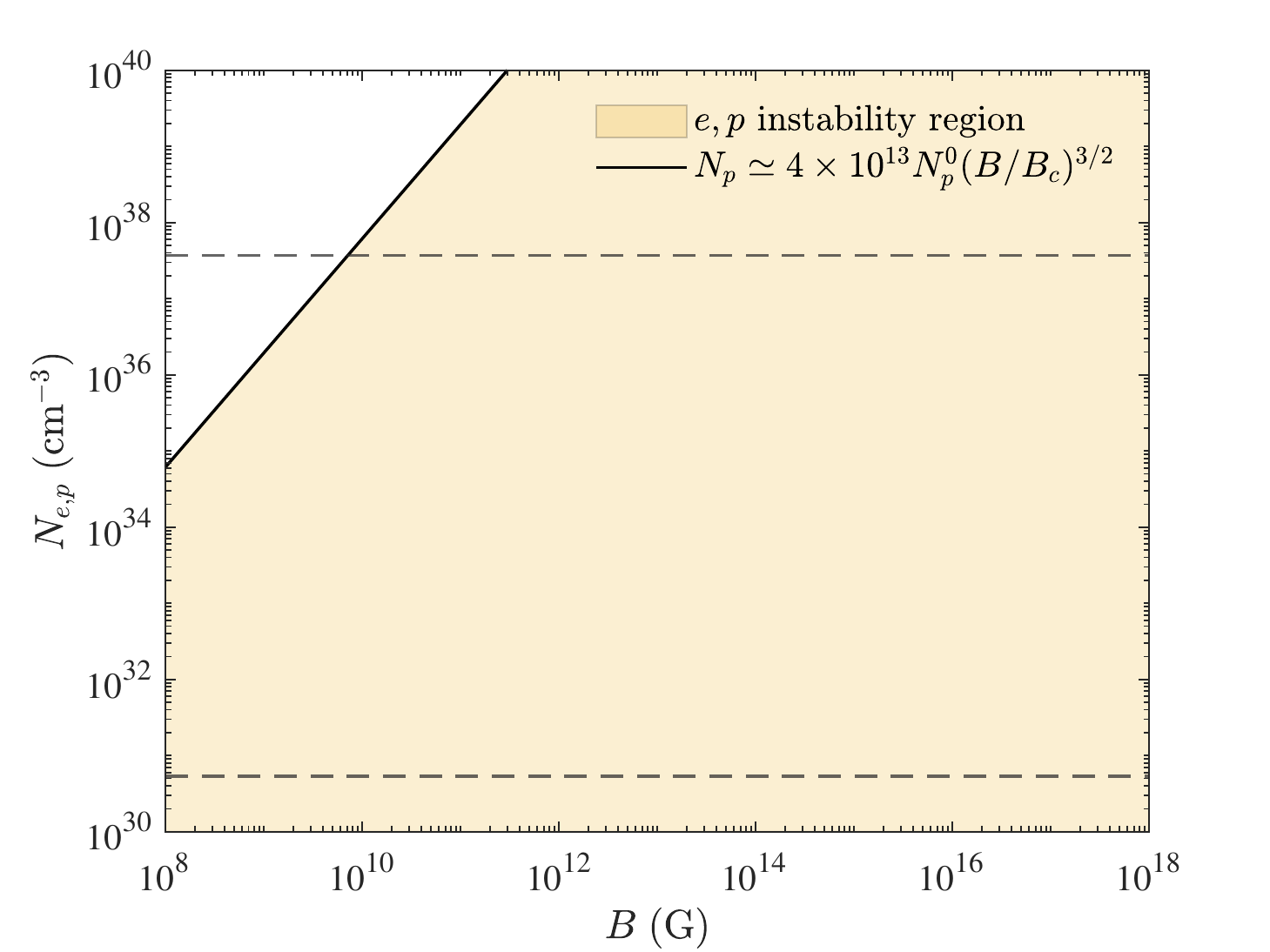}
\includegraphics[width=1\linewidth]{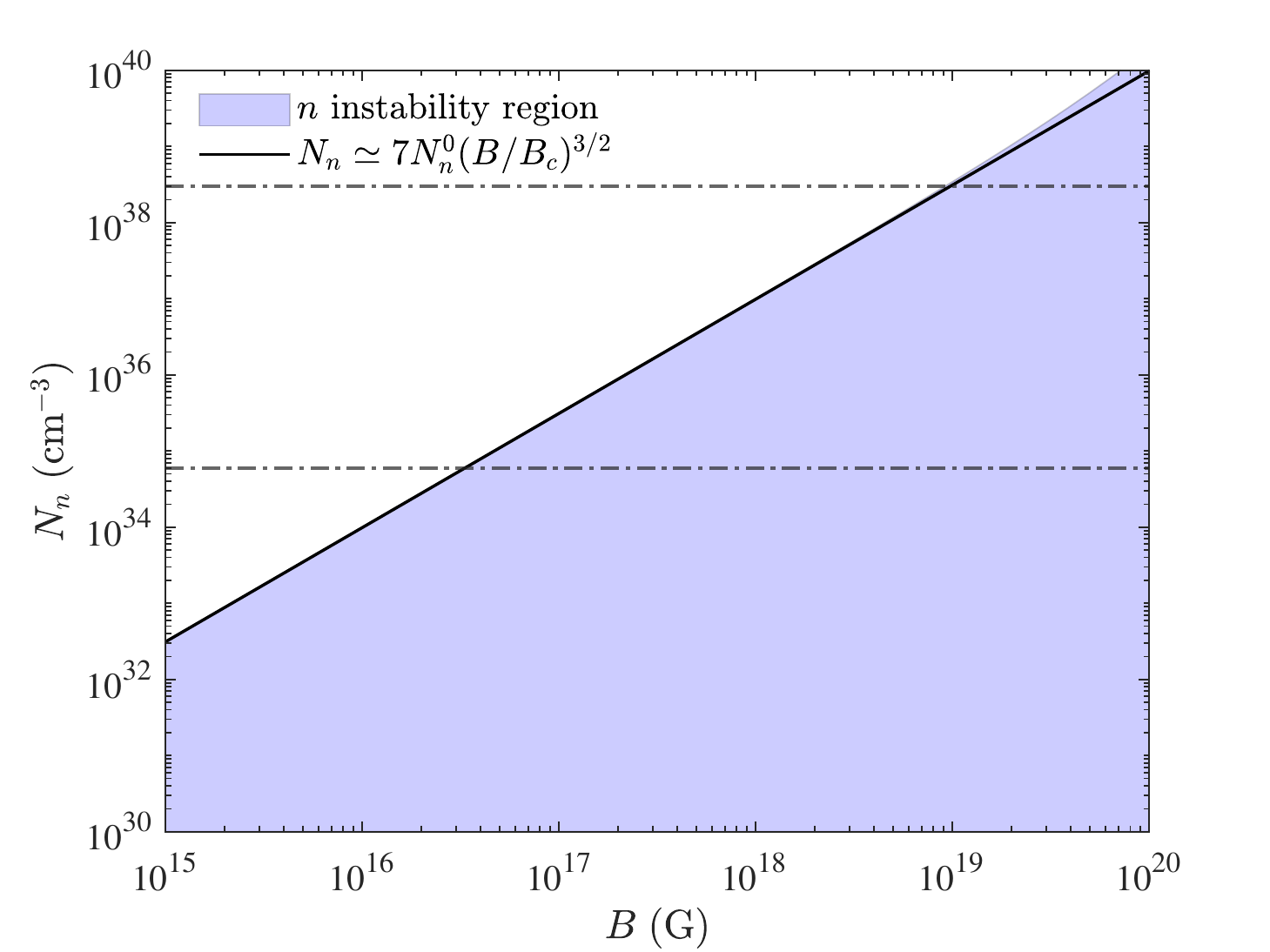}
\caption{\label{coll1} Phase diagrams for the transverse magnetic collapse of electrons and protons (upper plot), and neutrons (lower plot) inside a neutron star. In the colored region, $P_{\perp}(B,N)<0 $ and the gas is unstable. The horizontal lines delimit the typical ranges for the density of each particle inside the star.}
\end {figure}

In order to verify that the transverse magnetic collapse indeed occurs inside the NS, the following condition for the perpendicular pressure of at least one gas component must be satisfied:
\begin{equation}
P_{i\;\perp} (B,N_i) = -\Omega_i(B,N_i)-\mathcal M_i(B,N_i) B= 0,
\end{equation}
together with the stellar equilibrium conditions given in Eqs.~(\ref{estelarcond}). The results are illustrated in Fig.~\ref{coll1}. The solid lines correspond to $P_{i\;\perp} = 0$, which can be well approximated by the analytical fits $N_p \simeq 4\times10^{13} N_{p}^0 (B/B_c)^{3/2}$ and $N_n \simeq 7 N_{n}^0 (B/B_c )^{3/2}$ for protons and neutrons, respectively, with $N_{p}^0 \simeq N_{n}^0 \simeq 2.7\times10^{39}\,\rm cm^{-3}$ and the critical magnetic fields given by $B_c \simeq 1.49\times10^{20}$~G and $B_c \simeq 1.56\times10^{20}$~G for protons and neutrons, respectively. The colored area corresponds to the density and magnetic field values for which $P_{i \perp} < 0$ and the gas is unstable. Above the solid lines, $P_{i \perp} > 0$ and the gas is stable. The horizontal dashed lines delimit the approximate range for the typical densities of each particle species inside the NS.

The possibility of the magnetic collapse of the gas of electrons, protons and neutrons becomes evident in Fig.~\ref{coll1}. As can be seen in these plots, given a value of the magnetic field, there is a critical density for electrons, protons and neutrons below which the gas is unstable. This is due to the fact that the parallel pressure ($-\Omega$) of a Fermi system increases as the density increases, which helps to balance the ``magnetic pressure term" ($-\mathcal M B$) and stabilizes the gas. The critical densities of electrons and protons are of the order of those expected in NS and, in particular, for values of the magnetic field higher than $10^{10}$~G are always larger. In the case of neutrons, the critical densities are lower than those estimated for the crust and core of neutron stars, except when the magnetic field is very high ($B \gtrsim  3\times 10^{16}$~G).

\begin{figure}[t]
	\centering
	\includegraphics[width=1\linewidth]{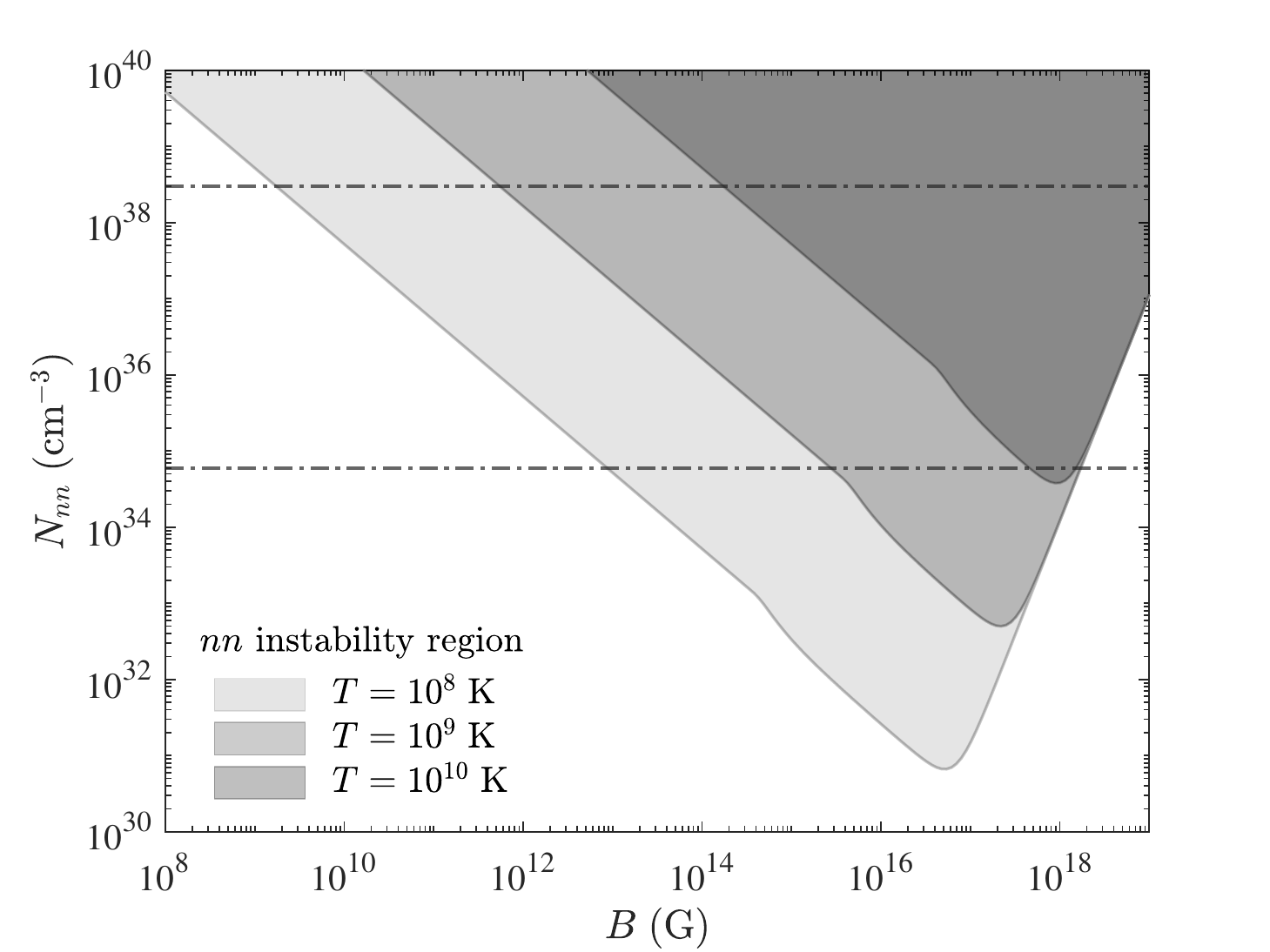}
	\caption{\label{coll2} Phase diagram for the transverse magnetic collapse of paired neutrons inside a neutron star. In the shaded gray regions, $P_{\perp}(B,N)<0 $ and the gas is unstable. The horizontal lines delimit typical ranges of neutron densities inside the star.}
\end {figure}

Unlike what happens for fermions, the critical density of paired neutrons is quite sensitive to the temperature, as can be observed in Fig.~\ref{coll2}, where the instability region is plotted for $T=10^{8}$~K, $T=10^{9}$~K, and $T=10^{10}$~K (from lighter to darker gray regions). Notice that the phase diagram is reversed in comparison to the case of fermions (cf. Fig.~\ref{coll1}), i.e., the gas becomes stable below the critical particle density. This occurs because the magnetization and magnetic pressure $-\mathcal M_i B$ increase as the density of a vector boson gas increases, enhancing the appearance of the Bose-Einstein condensate that diminishes the parallel pressure $-\Omega_i$. The sum of both effects is a decrease in the parallel pressure that for high enough densities makes the gas unstable. However, for magnetic fields $B\geq10^{17}$~G, a minimum appears and the critical curve starts increasing with $B$. At such strong magnetic fields the vacuum pressure term $-\Omega_{vac}$ becomes significant and helps to stabilize the gas. The critical densities of the paired neutrons are also in the range of those expected within a NS, in particular for $B\geq 10^{12}$~G, confirming also the possibility of the magnetic collapse of the paired neutron gas.

We have not shown the phase diagram for the paired protons. Unlike the other gases, the paired proton gas has a diamagnetic behavior, since the magnetization $\mathcal{M} =-eN/(2 \sqrt{m^2 + eB})<0$ (see Eq.~\eqref{EoSCSB4} of Appendix A). Consequently, the magnetic pressure term $-\mathcal M B$ is positive and it increases the perpendicular pressure, thus preventing the transverse magnetic collapse of this gas.

Let us summarize the results of this section. Under our assumptions, we have shown that a transverse magnetic collapse is possible for the gases of electrons, protons, neutrons and paired neutrons in a NS configuration. At this point, it is worth remarking that the conditions for such a collapse are assumed within the interior of the star. The NS is considered stable as a whole, although somewhere in its core some of the gases that compose it could become unstable. Therefore, in what follows, we assume that the densities and magnetic field values are such that the neutron gas component never collapses. We shall also consider that the conditions for the collapse of the remaining gases are local inside the star, or in other words, that the interior of the star is not homogeneous neither in density nor in the magnetic field.

\subsection{Ejection of matter from the neutron star}
\label{subsec2b}

In the previous section we have found that electrons, protons, and bosonized neutrons in the interior of a NS can produce an instability in the transverse pressure. However, the fact that these gases can collapse does not automatically imply that matter is ejected from the NS. For the ejection to take place, the collapsed gases must have a parallel pressure greater than the gravitational pressure of the star. This condition would in principle guarantee a way in which matter could escape from the star. 

For the study of the matter ejection mechanism described above, we shall compare the parallel pressures of the collapsed gases with some ``heuristic" value of the gravitational pressure. A rough estimate can be obtained considering the average gravitational pressure obtained for a compact object with a typical NS mass $M$ in the limit of compactness, $1.4M_{\odot} \lesssim M \lesssim 2.1 M_{\odot}$, where $M_{\odot}$ is the mass of the Sun.

The compactness solution corresponds to the analytical solution of the Tolman-Oppenheimer-Volkoff (TOV) equations, assuming that the inner density of the compact object is constant~\cite{Weinberg}. The radius and the pressure inside the star are given as functions of the total mass $M$, the radius $R$ of the star and its constant energy density $\mathcal{E}$~\cite{Weinberg},
\begin{align}\label{compactness}
R &= \left(\frac{3M}{4\pi\mathcal{E}}\right)^{1/3}, \\[2mm]
P(r) &= \mathcal{E}\frac{\sqrt {1-2 GM/R} -\sqrt{1-2 G M r^2/R^3}} {\sqrt{1-2 G M r^2 /R^3} -3\sqrt {1-2 G M/R}}\,,
\end{align}
where $G$ is the gravitational constant, $r$ is the internal radius of the star, and $\mathcal{E}$ is determined by the EoS of the partially bosonized $npe$-gas given in Eqs.~\eqref{edenpe}. The average pressure inside the star is defined as
\begin{equation}\label{Pmed}
P_{GAV}=\frac{1}{R}\int_{0}^{R}P(r)dr\,.
\end{equation}

\begin{figure}[t]
\centering
\includegraphics[width=1\linewidth]{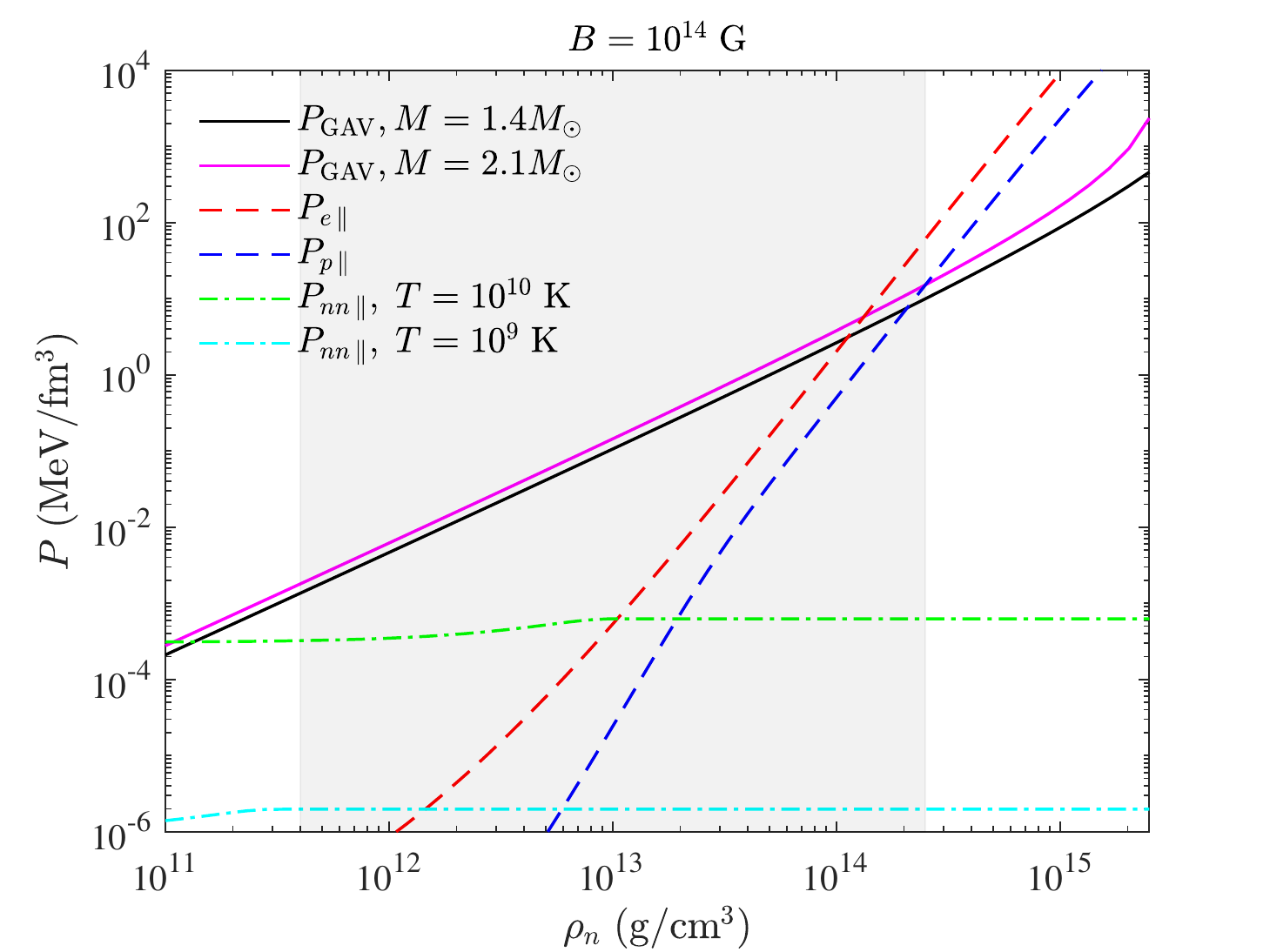}
\includegraphics[width=1\linewidth]{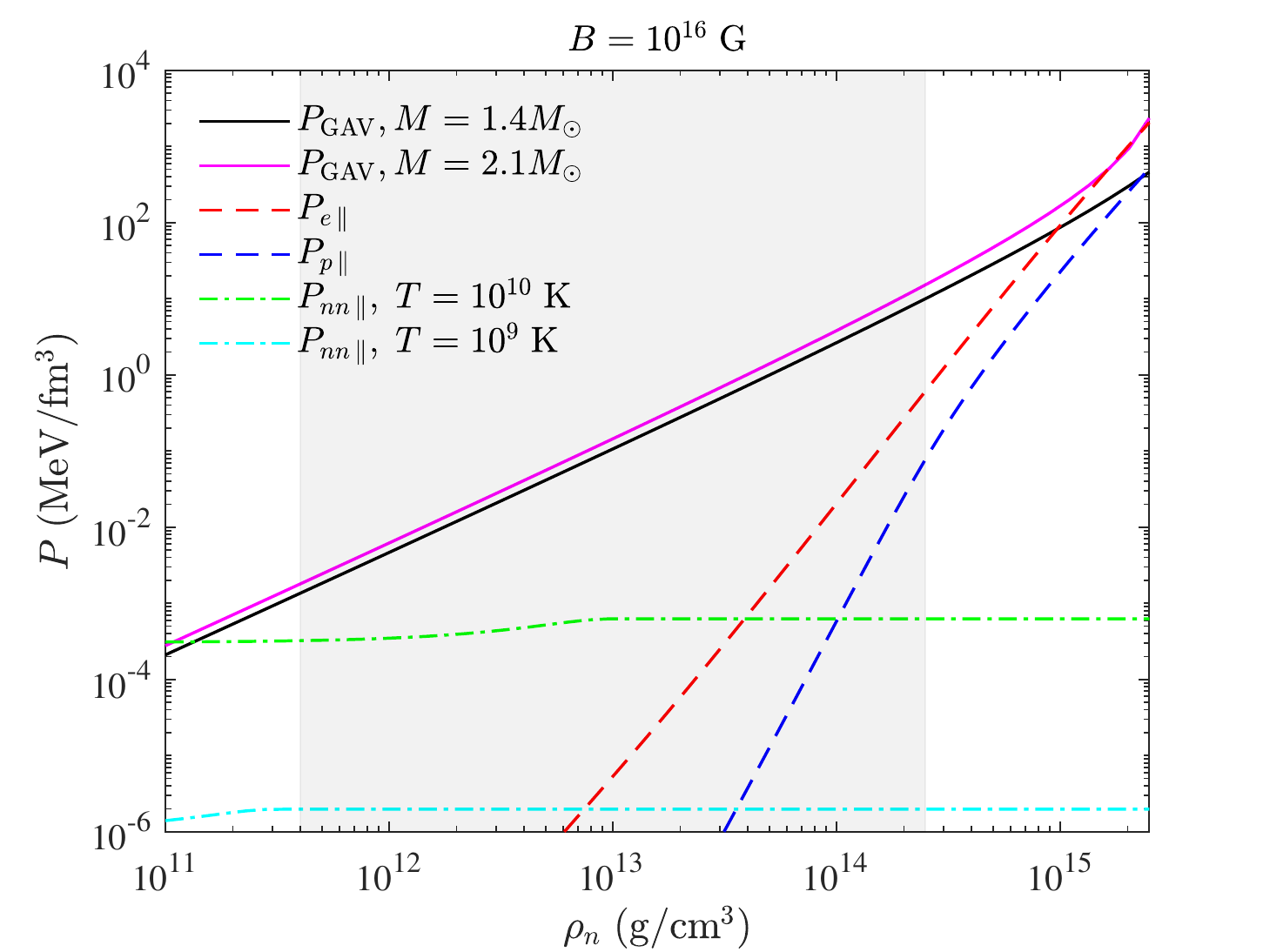}
\caption{\label{collapse3} Parallel pressure of electrons, protons and bosonized neutrons for typical neutron densities at the inner of a neutron star, for $B=10^{14}$~G and $B=10^{16}$~G. The shaded gray region delimits density values between the crust ($\rho \simeq 4 \times 10^{11}$~g/cm$^3$) and the core ($\rho \simeq 2.7 \times 10^{14}$~g/cm$^3$) of the star~\cite{Weber:2004kj}.}
\end{figure}

In Fig.~\ref{collapse3}, we show the comparison of $P_{GAV}$ with the parallel pressures of the electron, proton and bosonized neutron gases. For the magnetic fields and densities considered, the three gases have already collapsed. The neutron gas component is not included in this analysis because we assume that the stability of the NS is sustained by the non-negative neutron pressure. For values of the magnetic field around $10^{14}$~G, the average gravitational pressure can be overcome by the parallel pressures of the collapsed protons and electrons in the nucleus of the star ($\rho \sim 2.7 \times 10^{14}$~g/cm$^3$~\cite{Weber:2004kj}). As the magnetic field increases, the parallel pressures of the Fermi gases decrease. Instead, for the bosonized neutron gas, the parallel pressure remains almost constant over the range of magnetic field intensities, depending essentially on the temperature $T$, as can be seen in the figure. Nevertheless, for typical NS temperature and density ranges, $P_{nn\parallel}$ never exceeds $P_{GAV}$.

Therefore, depending on the values of the magnetic field and the particle density, the collapsed gases could overcome gravity and trigger the ejection of matter. In this sense, the origin of the jet will be different depending on the density and the magnetic field, that is, depending on the region of the NS from which it originates. Since the NS as a whole is considered a stable object, jet production will be set by local inhomogeneities of the magnetic field and density inside the star.

In the following, we will assume that the jet (or expelled matter from the star) has the same composition as the neutron star, that is, it will be also formed by a partially bosonized $npe$-gas, but in which there is at least one collapsing gas component.

\section{Self-generated magnetic field and collimated matter}
\label{sec3}

Once matter leaves the star and the jet is formed, a natural question arises: how does it remain collimated? The elongated shape of the jet could be easily explained if one or more gas components inside the star remain in a collapsed state, decreasing the perpendicular pressure and letting gravity to compress matter toward the jet axis. This is only possible if the magnetic field in the jet continues to be as intense as the one produced by the NS. 

From the observational point of view, it has been shown that the magnetic field of the jet is of the order of the magnetic field of its source~\cite{Carrasco2010_20,deGouveiaDalPino:2005xn}. Therefore, the question to be answered is whether matter can remain magnetized once it leaves the star. To address this question, we compute the magnetization of the partially bosonized $npe$-gas as a function of the magnetic field, and look for a self-generation of the magnetic field, i.e., for the existence of a solution of the equation $\mathcal{M}^{T} = B / 4\pi$.

\begin{figure}[t]
\centering
\includegraphics[width=1\linewidth]{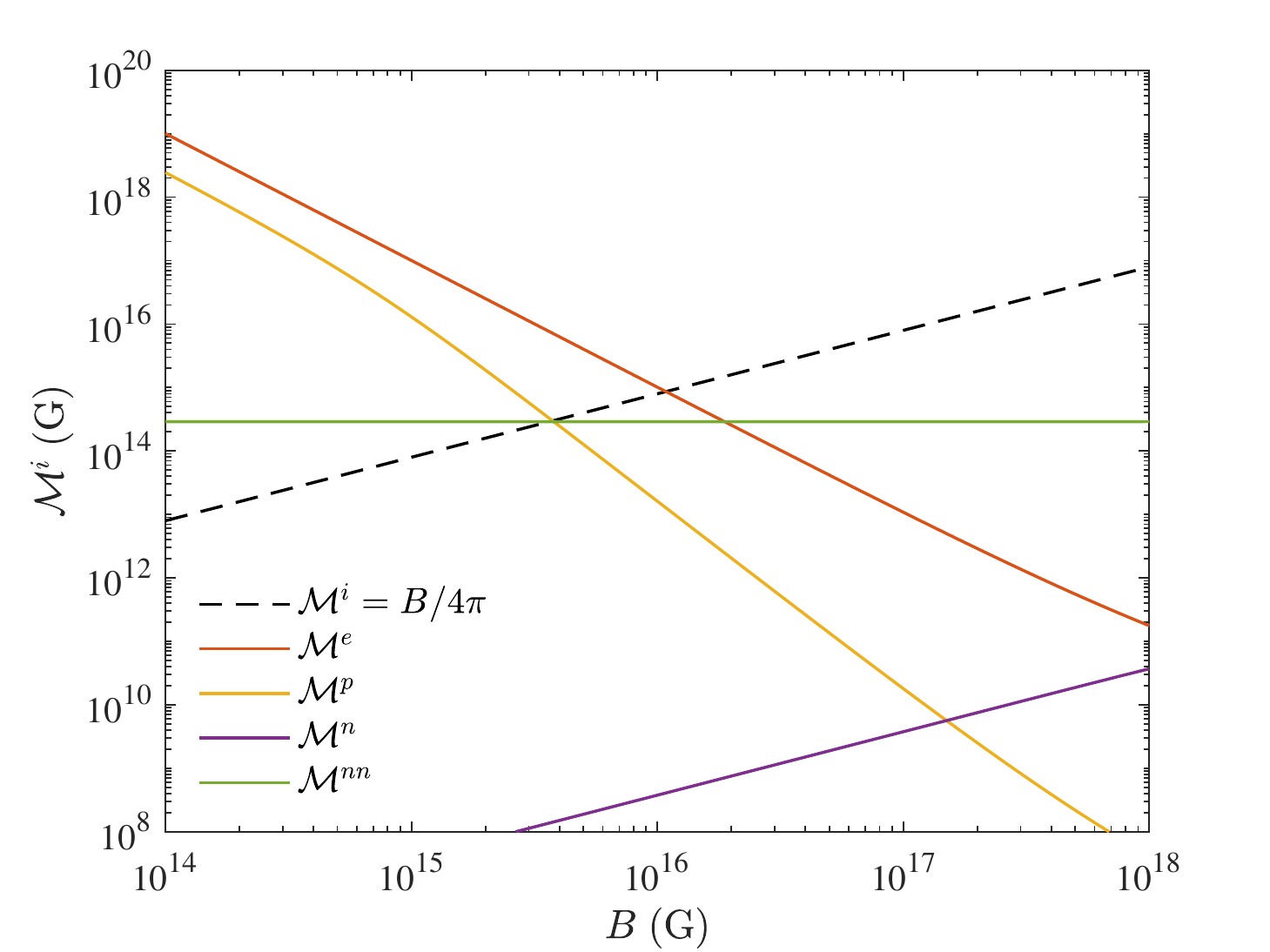}
\includegraphics[width=1\linewidth]{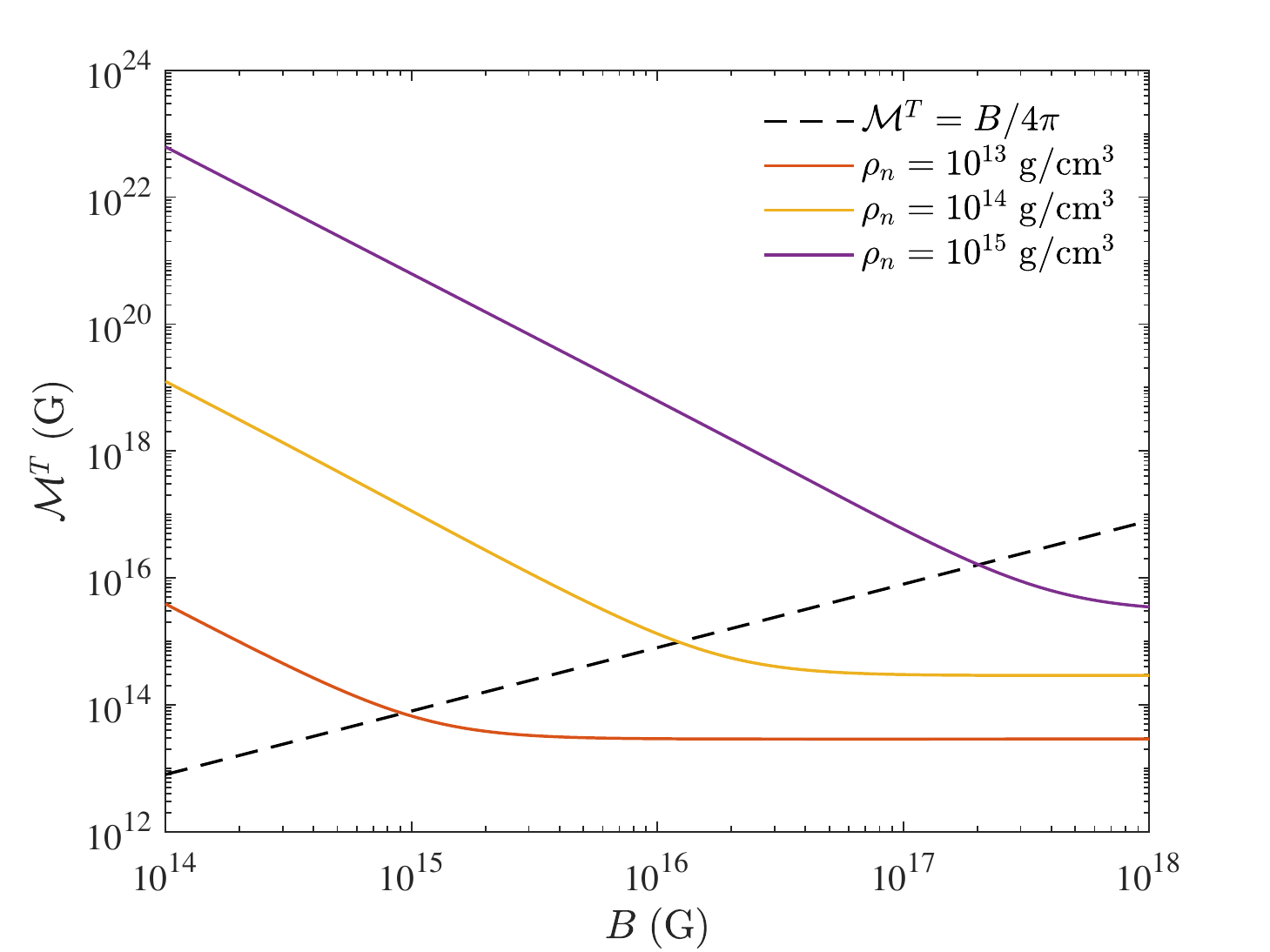}
\caption{\label{magnetizationnpe} Upper panel: the magnetization of the different components of the bosonized $npe$-gas as a function of $B$ for a neutron mass density $\rho_n=10^{14}$~g/cm$^3$. Lower panel: the total magnetization of the bosonized $npe$-gas as a function of $B$ for several values of the neutron mass density $\rho_n$.}
\end{figure}

\begin{figure}[t]
	\centering
	\includegraphics[width=1\linewidth]{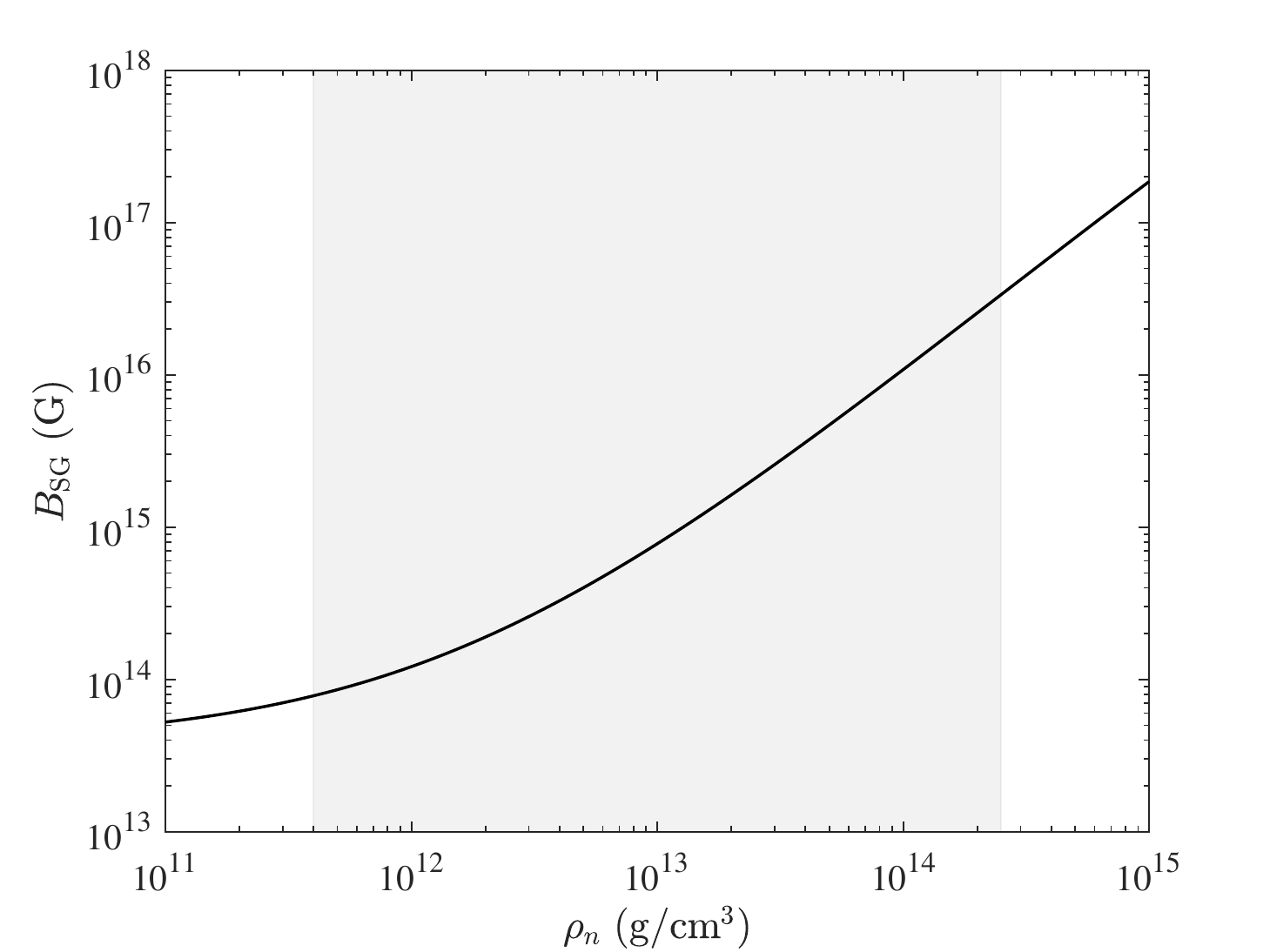}
	\caption{\label{selfmag} The self-generated magnetic field as a function of the neutron mass density. The shaded gray region corresponds to the typical density values between the crust and the core of the NS.}
\end{figure}

In the upper panel of Fig.~\ref{magnetizationnpe}, the magnetization of the different components of the bosonized $npe$-gas are plotted as functions of the magnetic field $B$ for a neutron mass density $\rho_n=10^{14}$~g/cm$^3$. As can be seen from this figure, for lower values of $B$ the magnetization is dominated by the electron and proton contributions, while for larger values of $B$ the magnetization of paired neutrons $\mathcal{M}^{nn}$ turns out to be the most significant. For the range of magnetic field values considered, the latter quantity remains almost constant and is given by $\mathcal{M}^{nn}\simeq \kappa_{nn} \rho_n/(4m_n) \simeq 2.9\times10^{14}$~G. 

In the lower panel of Fig.~\ref{magnetizationnpe}, we present the total magnetization of the $npe$-gas for several values of the neutron mass density. The dashed line corresponds to $B/4\pi$ and it intersects the curves indicating that the self-magnetization condition might be fulfilled. This is shown in Fig.~\ref{selfmag}, in which the self-generated magnetic field $B_ {SG}$ is plotted as a function of the neutron mass density. We notice that, depending on the density, the values of the self-generated magnetic field can be high enough ($B_ {SG}\geq 10^{13}$~G) to keep the gases that form the jet in the collapsed regime. In the region of the typical density values between the crust and the core of the NS, the self-generated magnetic field is $8\times 10^{13}~\text{G} \lesssim B_ {SG} \lesssim 4\times 10^{16}$~G.

\begin{figure}[t]
	\centering
	\includegraphics[width=1\linewidth]{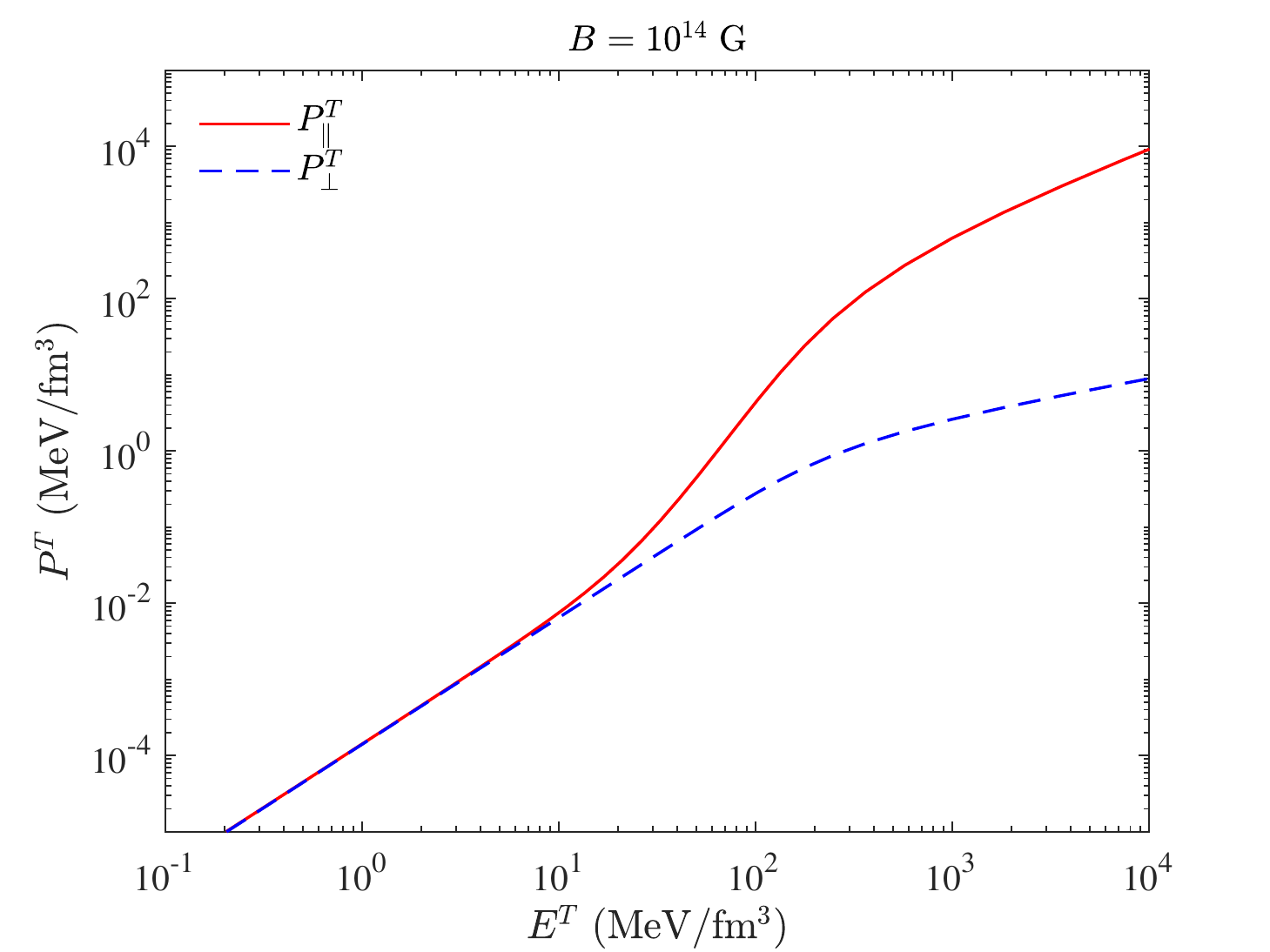}
	\includegraphics[width=1\linewidth]{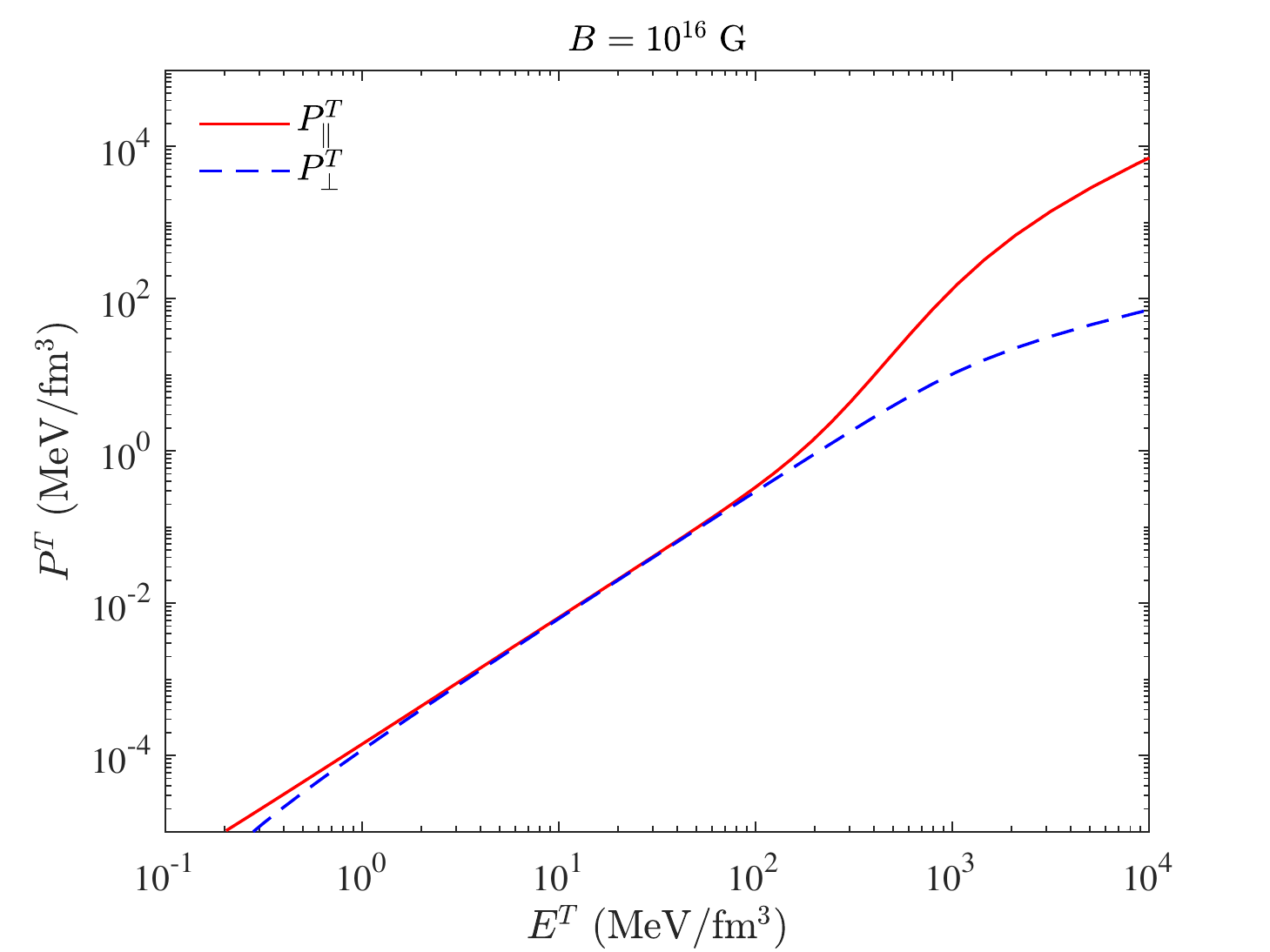}
	\caption{\label{pressures} The total parallel and perpendicular pressures of the jet as a function of its internal energy for $B=10^{14}$~G and $B=10^{16}$~G.}
	\end {figure}

We have established that the transverse magnetic collapse is a viable mechanism for the ejection of matter from a NS, and that the expelled matter can produce a magnetic field high enough to keep it collimated. Let us now proceed to characterize the EoS of the jet through the pressure dependence on the internal energy density and the magnetic field intensity. This is shown in Fig.~\ref{pressures}, where the total parallel and perpendicular pressures of the jet are presented as functions of the internal energy for $B=10^{14}$~G and $B=10^{16}$~G. As before, we assume an equal fraction of bosonized baryons, i.e., $x_n=x_p=0.5$. We note that the difference between the parallel and perpendicular pressures of the jet can reach up to three orders of magnitude in the region of high energy densities, being this difference a function of $B$ (and $x_n$). For lower densities, both pressures are approximately equal. From the gravitational equilibrium viewpoint, the difference in the pressures is related to a difference in the jet dimensions~\cite{Lattimer:2000nx} that could account for its elongated form. Therefore, this EoS could describe an axially symmetric structure of the jet as far as its interior densities are in the region where the pressures are anisotropic.

\section{Concluding remarks}
\label{sec4}

In this work, we have studied the transversal magnetic collapse of matter inside a neutron star, considering it as a partially bosonized $npe$-gas. We have found that, in typical conditions for the NS interior, the electron, proton, neutron, and paired neutron gases are susceptible to suffer a transverse magnetic collapse, while the gas of paired protons is always stable. 

Under certain conditions of the magnetic field strength and particle densities, the collapsed gases could overcome gravity and trigger the expulsion of matter out of the star. Moreover, the collapsed matter can be self-magnetized, producing a magnetic field strong enough to keep the gases in a collapse configuration once they leave the star. Since the difference between the parallel and perpendicular pressures of the collapsed matter can reach up to three orders of magnitude, the EoS obtained in this framework is expected to describe a highly elongated structure such as a jet.

The mechanism advocated here for astrophysical jet production and maintenance is based on the properties of strongly magnetized quantum gases. The EoS was studied assuming that the source is a neutron star composed of a partially bosonized $npe$-gas, in which at least one of the gas species in the mixture collapses. Nevertheless, the proposed mechanism could also be applied to other compact objects as long as they contain gases susceptible to suffer a transversal magnetic collapse. Therefore, it could be useful for explaining the physical phenomenon of matter ejection in other astrophysical environments.

Finally, an important issue to be addressed is the gravitational stability of the jet. The search for stable solutions of Einstein's equations with an anisotropic EoS of matter requires numerical methods to integrate Einstein's equations in an axisymmetric space-time. This is a nontrivial problem that is still unsolved. Some attempts at solving this problem have been carried out in the context of other physical situations. Yet, those results cannot be applied to the description of such elongated structures as astrophysical jets. For instance, in Ref.~\cite{AlvearTerrero:2019aqi}, a search for inner solutions of Einstein's equations assuming an almost flat-symmetric metric was performed, since this type of space-time adapts to the symmetry that the magnetic field imposes on the energy-momentum tensor. However, the use of this type of space-time imposes conditions to the pressures and energy that are only fulfilled in very restrictive situations (e.g., when the pressures and energy of the vacuum dominate over those of matter). In another attempt, structure equations for spheroidal objects were obtained~\cite{Terrero:2018utx}. The latter equations are restricted to the case of small deformations with respect to the spherical shape and are not appropriate to describe a jet. The gravitational stability of jets is certainly an interesting problem that deserves further study.

\section{Acknowledgments}
The authors thank Gabriel Gil P\'erez for useful comments. G.A.Q and A.P.M have been supported by the grant No.~500.03401 of the PNCB-MES, Cuba, and the grant of the Office of External Activities of the Abdus Salam International Centre for Theoretical Physics (ICTP) through NT-09. A.P.M. expresses her gratitude to Fermilab and ICTP for their hospitality. The work of R.G.F. was partially supported by Funda\c{c}{\~a}o para a Ci{\^e}ncia e a Tecnologia (FCT, Portugal) through the projects CFTP-FCT Unit 777 (UID/FIS/00777/2019), CERN/FIS-PAR/0004/2017 and PTDC/FIS-PAR/29436/2017, which are partly funded through POCTI (FEDER), COMPETE, QREN and EU.

\appendix

\section{Equations of state of magnetized quantum gases}\label{appA}

In this appendix, we present the equations of state of charged and neutral magnetized gases of fermions and bosons used in our calculations. These are obtained considering a uniform and constant magnetic field in the $z$ direction, $\textbf{B}=(0,0,B)$. 

Hereafter, $m$ is the particle mass, $q$ its charge, $\kappa$ the magnetic moment of neutral particles, $B_c$ is the critical magnetic field at which the magnetic energy of the particle becomes comparable to its rest mass, and $b \equiv B/B_c$.

\subsection{Charged fermions}

The EoS for a gas of charged fermions (electrons and protons) at low temperatures are~\cite{Felipe:2002wt}
\begin{subequations}\label{EoSCF}
	\begin{align}
 E&= \frac{m^2 b}{4\pi^2}\sum_{l=0}^{l_{max}} g_{l}\! \left[ \mu\,p_F +{\mathcal {E}_{l}}^2\ln\left(\frac{ {\mu}+ {p_F}}{\mathcal {E}_{l}}\right)\right],\\[2mm]
  \label{EoSCF2}
  P_{\parallel}
  &= \frac{m^2 b}{4\pi^2}\sum_{l=0}^{l_{max}}g_{l}\!\left[ \mu\,p_F -{\mathcal {E}_{l}}^2\ln\!\left(\frac{\mu+ p_F}{\mathcal {E}_{l}}\right)\right]\!,\\[2mm]
  \label{EoSCF4}
  P_{\perp} &=\frac{m^4 b^2}{2\pi^2}\,\sum_{l=0}^{l_{max}}g_l\, l\,\ln\left (\frac{\mu+p_F}{\mathcal{E}_l}\right),\\[2mm]
  \label{EoSCF5}
  \mathcal M &= \frac{m^2}{4\pi^2 B_c} \sum_{l=0}^{l_{max}}g_{l}\!\left[ \mu \,p_F - \mathcal {E}_{l}^{2} \ln \left(\frac{\mu + p_F}{\mathcal {E}_{l}}\right) \right]\!  ,\\[2mm]
  \label{EoSCF6}
  N &= \frac{m^2 b}{4\pi^2} \sum_{l=0}^{l_{max}} 2\, g_{l}\, p_F,
	\end{align}
	\end{subequations}
where $E$ denotes the internal energy, $P_{\parallel}$ and $P_{\perp}$ are the pressure components along and perpendicular to the magnetic field direction, respectively, $\mathcal M$ is the magnetization, and $N$ is the particle number density.

In Eqs.~\eqref{EoSCF}, the integer numbers $l$ are the Landau levels, $g_{l}=2-\delta_{0,l}$, $l_{max}= I[(\mu^2-m^2)/(2qB)]$ and $I[z]$ denotes the integer part of $z$. The Fermi momentum is $p_F=(\mu^2-\mathcal{E}_{l}^2)^{1/2}$, with the rest energy given as $\mathcal{E}_{l}=(2 qB l+m^2)^{1/2}$ and the quantity $B_c = m^2/q$.

\subsection{Neutral fermions}

For a gas of neutral fermions (neutrons) the EoS read as~\cite{Aurora2003EPJC}
\begin{subequations}
\begin{align}
  E &=- P_{\parallel} + \mu N, \label{EoSNF2}\\[2mm]
  P_{\parallel} &=  \frac{m^4}{2\pi^2} \sum_{\eta=1,-1} \left[ \frac{\mu f^3}{12 m} + \frac{(1+\eta b)(5 \eta b -3)\mu f}{24 m} \right.\nonumber\\
   &+ \left.\frac{(1+\eta b)^3 L}{24} -\frac{\eta b \mu^3 s}{6 m^3} \right],  \label{EoSNF3}\\[2mm]
  P_{\perp} &= P_{\parallel} - \mathcal M B,  \label{EoSNF4}\\[2mm]
  \mathcal M &= \frac{m^3 \kappa}{2\pi^2} \sum_{\eta=1,-1} \eta \left[ \frac{(1-2 \eta b)}{6} f 
  \right.\nonumber\\
  &+ \left. \frac{(1 + \eta b)^2(1-\eta b/2)}{3} L - \frac{\mu^3}{6 m^2} s \right],  \label{EoSNF5}\\[2mm]
N &= \frac{m^3}{2\pi^2} \sum_{\eta=1,-1} \left[ \frac{f^3}{3} + \frac{\eta b(1+\eta b)}{2}f - \frac{\eta b \mu^2}{2 m^2}s  \right], \label{EoSNF1}
\end{align}
\end{subequations}
where $\eta$ denotes the spin states, $B_c=m/\kappa$ and the functions $f$, $L$ and $s$ are defined as
\begin{eqnarray}
  f &=& \frac{\sqrt{\mu^2-\varepsilon(\eta)^2}}{m},\\
  L &=& \frac{1}{1+\eta b} \ln\left(\frac{\mu + \sqrt{\mu^2 - \varepsilon(\eta)^2}}{m}\right),\\
  s &=& \frac{\pi}{2}- \arcsin\left[\frac{m}{\mu}\left(1+\eta b\right)\right],
\end{eqnarray}
with $\varepsilon(\eta) = m + \eta\, \kappa\, B$.

\subsection{Charged scalar bosons}

In this section, we present the EoS for a gas of charged scalar bosons (paired protons),\footnote{To our knowledge, these equations have not been previously presented in the literature.} following the same procedure developed in Refs.~\cite{Khalilov:1997xu} and \cite{Quintero2017PRC}. We have  checked that our results are consistent with those obtained in the very low temperature limit~\cite{ROJAS1996148}.

For a charged scalar gas of bosons, the EoS read in the weak-field (WF) and strong-field (SF) regimes as
\begin{subequations}\label{EoSCSB}
	\begin{align}\label{EoSCSB1}
  E &= \left \{
        \begin{array}{ll}
        \varepsilon N - 3\Omega/2, & \, {\rm WF}, \\[2mm]
        \varepsilon N - \Omega/2, & \, {\rm SF},
         \end{array}
         \right.\\[2mm]
  \label{EoSCSB2}
  P_{\parallel} &= -\Omega ,\\[2mm]
  \label{EoSCSB3}
  P_{\perp} &= P_{\parallel} - \mathcal M B ,\\[2mm]
  \label{EoSCSB4}
 \mathcal M &=  - \frac{e}{2 \varepsilon} N,\\[2mm]
  \label{EoSCSB5}
  N &=\left \{
        \begin{array}{ll}
        \dfrac{3 \varepsilon^{3/2} Li_{3/2}(e^{\beta\mu^{\prime}})}{(2 \pi \beta)^{3/2}}, & \, {\rm WF},  \\[5mm]
        \dfrac{3 m^2 b\, \varepsilon^{1/2} Li_{1/2}(e^{\beta\mu^{\prime}})}{(2 \pi)^{3/2} \beta^{1/2}}, & \, {\rm SF}.
         \end{array}
         \right.
         	\end{align}
\end{subequations}

In Eqs.~\eqref{EoSCSB}, $\mu^{\prime} = \mu - \varepsilon$, $\varepsilon = 2 m_{p} \sqrt{1+b}$, $B_c = 2 m_{p}^2/e$ and $Li_{k}(x)$ is the polylogarithmic function of order $k$. The weak ($T > m_p b$)  and strong ($T < m_{p} b$) field regimes are separated by the condition $T = m_{p} b$. Furthermore,
\begin{equation}
\centering
 \Omega = \left \{
        \begin{array}{lr}
        -\dfrac{3 \varepsilon^{3/2} Li_{5/2}(e^{\beta(\mu-\varepsilon)})}{(2 \pi)^{3/2} \beta^{5/2}} , & \, {\rm WF}, \\[5 mm]
        -\dfrac{3 m^2 b \varepsilon^{1/2} Li_{3/2}(e^{\beta(\mu-\varepsilon)})}{(2 \pi)^{3/2} \beta^{3/2}}, & \, {\rm SF},
         \end{array}\right.
	\end{equation}
where
\begin{equation}
 \mu = \left \{
        \begin{array}{ll}
        \varepsilon-\dfrac{\zeta(3/2) T}{4 \pi} \left[ 1-\left(\dfrac{T_{c}}{T}\right)^{3/2}\right] \Theta(T - T_{c}), & \, {\rm WF}, \\[5mm]
        \varepsilon \left( 1- \dfrac{m^4 b^2 T^2}{8 \pi^2 N^2}\right), & \, {\rm SF},
         \end{array}
         \right.
\end{equation}
and
\begin{equation}
 T_{c} = \frac{2 \pi}{\varepsilon} \left[ \frac{N}{3\, \zeta(3/2)}\right]^{2/3}
\end{equation}
is the condensation temperature and $\zeta(x)$ is the Riemann zeta function.

We conclude from Eq.~\eqref{EoSCSB4} that the magnetization $\mathcal M$ is always negative for a gas of charged scalar bosons and, therefore, a magnetized charged boson gas never collapses magnetically.

\subsection{Neutral vector bosons}

For a gas of neutral vector bosons (paired neutrons) in the limit of low temperatures the EoS read~\cite{Quintero2017PRC}
\begin{subequations}
\begin{align}\label{EoSNB1}
  E &= \varepsilon N + \Omega_{vac}- \frac{3}{2} \Omega_{st} - \frac{\partial \mu^{\prime}}{\partial \beta} N,\\[2mm]
  P_{\parallel} &= -\Omega_{st} -\Omega_{vac},\\[2mm]
  P_{\perp} &= -\Omega_{st} -\Omega_{vac}-\mathcal M B,\\[2mm]
  \mathcal M&= \frac{\kappa}{\sqrt{1-b}} N,\\[2mm]
  N&= N_c +\frac{\varepsilon^{3/2} Li_{3/2}(e^{\mu^\prime \beta})}{\sqrt{2\pi}\, \pi \beta^{3/2} (2-b)},
\end{align}
\end{subequations}
with $\beta =1/T$ the inverse temperature, $\mu^{\prime} = \mu - \varepsilon$, $\varepsilon = m_{n} \sqrt{1-b}$ the rest energy, $B_c = m_n/(2 \kappa)$, and $N_c=N\left[1-(T/T_c)^{3/2}\right]$ the density of condensed particles. Moreover,
\begin{equation}
\Omega_{st}=-\dfrac{\varepsilon^{3/2}Li_{5/2}(e^{\beta \mu^{\prime}})}{\sqrt{2\pi}\, \pi \beta^{5/2} (2-b)}
\end{equation}
is the statistical contribution to the thermodynamical potential,
\begin{eqnarray}
\Omega_{vac}&=&-\dfrac{m^4}{288 \pi} \bigg [ b^2(66-5 b^2)-3(6-2b-b^2) \bigg.\nonumber\\
&&\bigg. \times(1-b)^2 \log(1-b) -3(6+2b-b^2) \bigg.\nonumber\\
&&\bigg. \times(1+b)^2 \log(1+b) \bigg ]
\end{eqnarray}
corresponds to the vacuum term, and
\begin{eqnarray}
\mu^{\prime} &=& -\frac{\zeta(3/2)T}{4 \pi} \left [ 1- \left(\frac{T_{c}}{T} \right)^{3/2} \right ]\Theta(T-T_{c}),\\[2mm]
T_{c} &=& \frac{1}{\varepsilon} \left [ \frac{\sqrt{2\pi}\, \pi (2-b) N}{\zeta(3/2)}\right]^{2/3}.
\end{eqnarray}


\begin{thebibliography}{99}

\bibitem{Abbott:2016blz} 
  B.~P.~Abbott {\it et al.} [LIGO Scientific and Virgo Collaborations],
  Phys.\ Rev.\ Lett.\  {\bf 116}, 061102 (2016).

\bibitem{TheLIGOScientific:2017qsa} 
  B.~P.~Abbott {\it et al.} [LIGO Scientific and Virgo Collaborations],
  Phys.\ Rev.\ Lett.\  {\bf 119}, 161101 (2017).
  
\bibitem{GBM:2017lvd}
  B.~P.~Abbott {\it et al.},
  Astrophys.\ J.\  {\bf 848}, L12 (2017).

\bibitem{deGouveiaDalPino:2004jy}
  E.~M.~de Gouveia Dal Pino,
  Adv.\ Space Res.\  {\bf 35}, 908 (2005).

\bibitem{Blandford:2018iot} 
  R.~Blandford, D.~Meier and A.~Readhead,
  Annual Review of Astronomy and Astrophysics {\bf 57}, 467 (2019).
 
\bibitem{deGouveiaDalPino:2005xn}
  E.~M.~de Gouveia Dal Pino,
  AIP Conf.\ Proc.\  {\bf 784}, 183 (2005).

\bibitem{Pino:2010qs} 
  E.~M.~d.~Gouveia Dal Pino, P.~P.~Piovezan and L.~H.~S.~Kadowaki,
  Astron.\ Astrophys.\  {\bf 518}, A5 (2010).

\bibitem{Tucker:2016wvt}
  R.~W.~Tucker and T.~J.~Walton,
  Class.\ Quant.\ Grav.\  {\bf 34}, 035005 (2017).

\bibitem{Bini:2007zzb} 
  D.~Bini, F.~de Felice and A.~Geralico,
  Phys.\ Rev.\ D {\bf 76}, 047502 (2007).

\bibitem{Chicone:2010aa} 
  C.~Chicone and B.~Mashhoon,
  Phys.\ Rev.\ D {\bf 83}, 064013 (2011).

\bibitem{Bini:2017uax}
D.~Bini, C.~Chicone and B.~Mashhoon,
Phys.\ Rev.\ D {\bf 95}, 104029 (2017).

\bibitem{Poirier:2015hga} 
  J.~Poirier and G.~J.~Mathews,
  arXiv:1504.02499 [gr-qc].

\bibitem{Massi:2010gr}
M.~Massi,
Mem.\ Soc.\ Ast.\ It.\  {\bf 82}, 24 (2011).

\bibitem{Lebedev:2001fa}
  S.~V.~Lebedev {\it et al.},
  Astrophys.\ J.\  {\bf 564}, L113 (2002).

\bibitem{Hartigan} Hartigan P., Astrophys Space Sci (2005) 298: 99.

\bibitem{Meier2012}
 D. L.~Meier, \emph{Black Hole Astrophysics: The Engine Paradigm}, Berlin: Springer-Verlag, 2012.

\bibitem{Nakamura:2018htq} 
  M.~Nakamura {\it et al.},
  Astrophys.\ J.\  {\bf 868}, 146 (2018).
 
\bibitem{Armengol:2016xhu} 
  F.~G.~Lopez Armengol and G.~E.~Romero,
  Astrophys.\ Space Sci.\  {\bf 362}, 214 (2017).

\bibitem{2000PhRvL..84.5261C} M. Chaichian, S. S. Masood, C. Montonen, A. P{\'e}rez Mart{\'\i}nez and H. P{\'e}rez Rojas, Phys. Rev. Lett. {\bf 84}, 5261 (2000).

\bibitem{Ferrer:2015wca}  E. J. Ferrer, V. de la Incera, D. Manreza--Paret, A. P\'erez--Mart\'{\i}nez and A. S\'anchez, Phys. Rev. D {\bf 91}, 085041 (2015).

\bibitem{Felipe:2008cm}
R.~G.~Felipe and A.~P.~Martinez, J.\ Phys.\ G {\bf 36}, 075202 (2009).

\bibitem{Felipe:2002wt} R. G. Felipe, H. J. Mosquera Cuesta, A. P\'erez Mart\'{\i}nez and H. P\'erez Rojas, Chin. J. Astron. Astrophys., {\bf 5}, 399 (2005).

\bibitem{Aurora2003EPJC} A. P{\'e}rez Mart{\'{\i}}nez, H. P{\'e}rez Rojas, H. J. Mosquera Cuesta, Eur. Phys. J. C, {\bf 29}, 111 (2003).

\bibitem{Quintero2017PRC}
G.~Q. Angulo, A.~Perez~Martinez, and H.~P. Rojas, 
 Phys. Rev. C {\bf 96}, 045810 (2017).
	
\bibitem{Chavanis}
P.~H.~Chavanis and T.~Harko,
Phys.\ Rev.\ D {\bf 86}, 064011 (2012).

\bibitem{Angulo:2018url} 
  G.~Quintero Angulo, A.~Pérez Martínez, H.~Pérez Rojas and D.~Manreza Paret,
  Int.\ J.\ Mod.\ Phys.\ D {\bf 28}, 1950135 (2019).

\bibitem{ROJAS1996148} H.~Rojas, Phys. Lett. B {\bf 379}, 148 (1996).

\bibitem{Weinberg}
S.~Weinberg, {\em {Gravitation and Cosmology: Principles and Applications of the General Theory of Relativity}}, New York, NY: Wiley, 1972.

\bibitem{Shapiro}
S.~L. {Shapiro} and S.~A. {Teukolsky}, {\em {Black holes, white dwarfs, and neutron stars: The physics of compact objects}}, New York, NY: Wiley, 1983.

\bibitem{Weber:2004kj} 
F.~Weber,
Prog.\ Part.\ Nucl.\ Phys.\  {\bf 54}, 193 (2005).

\bibitem{Carrasco2010_20}
C.~Carrasco-Gonz{\'a}lez, L.~F. Rodr{\'\i}guez, G.~Anglada, J.~Mart{\'\i}, J.~M. Torrelles, and M.~Osorio, 
Science~{\bf 330}, 1209 (2010).

\bibitem{Lattimer:2000nx} 
  J.~M.~Lattimer and M.~Prakash,
  Astrophys.\ J.\  {\bf 550}, 426 (2001).

\bibitem{AlvearTerrero:2019aqi} 
  D. Alvear Terrero, P. Bargue\~no, E. Contreras, A. P. Mart\'{\i}nez and G. Q. Angulo,
  Int.\ J.\ Mod.\ Phys.\ D {\bf 28}, 1950090 (2019).

\bibitem{Terrero:2018utx} 
  D. A. Alvear Terrero, V. Hern\'andez Mederos, S. L\'opez P\'erez, D. Manreza Paret, A. P\'erez Mart\'{\i}nez and G. Quintero Angulo,
  Phys.\ Rev.\ D {\bf 99}, 023011 (2019).

\bibitem{Khalilov:1997xu} 
  V.~R.~Khalilov, C.~L.~Ho and C.~Yang,
  Mod.\ Phys.\ Lett.\ A {\bf 12}, 1973 (1997).

\end{thebibliography}
\end{document}